\documentclass[aps,prb,twocolumn,floatfix,superscriptaddress,showpacs,amsmath,amssymb]{revtex4}
\usepackage{graphicx}
\usepackage{epsfig} 
\allowdisplaybreaks

\newcommand{\be}{\begin{equation}}
\newcommand{\ee}{\end{equation}}
\newcommand{\bea}{\begin{eqnarray}}
\newcommand{\eea}{\end{eqnarray}}
\newcommand{\ba}{\begin{eqnarray*}}
\newcommand{\ea}{\end{eqnarray*}}

\newcommand{\m}[1]{\mathcal{#1}} 
\newcommand{\eps}{\varepsilon}

\begin{document}

\title{Tunable Hybrid Quantum Electrodynamics from Non-Linear Electron Transport}
\author{Marco Schir\'o}
\affiliation{Department of Chemistry, Columbia University, New York, New York 10027, U.S.A.}
\affiliation{Institut de Physique Th\'{e}orique, CEA, CNRS-URA 2306, F-91191, 
Gif-sur-Yvette, France}
\author{Karyn Le Hur}
\affiliation{Centre de Physique Théorique (CPHT), École Polytechnique, CNRS, 91128 Palaiseau Cédex, France} 
  
\date{\today} 
 
\pacs{73.23.-b,42.50.-p,72.10.Fk}

\begin{abstract} 
Recent advances in quantum electronics have allowed to engineer hybrid nano-devices comprising on chip a microwave electromagnetic resonator coupled to an artificial atom, a quantum dot. These systems realize novel platforms to explore non-equilibrium quantum impurity physics with light and matter. Coupling the quantum dot system to reservoir leads (source and drain) produces an electronic current as well as dissipation when applying a bias voltage across the system. Focusing on a standard model of biased quantum dot coupled to a photon mode which gives rise to an Anderson-Holstein Hamiltonian, we elucidate the signatures of the electronic correlations in the phase of the transmitted microwave signal. In addition, we illustrate the effect of the electronic degrees of freedom on the photon field, giving rise to non-linearities, damping and dissipation, and discuss how to control these effects by means of gate and bias voltages. 
\end{abstract}
\maketitle

\section{Introduction} 
Recent years have seen tremendous experimental progress in probing and controlling quantum mechanics at the level of single constituents. A variety of physical systems, at the interface between quantum optics, atomic and solid state physics, have emerged as natural platforms to store and process information at the genuine quantum level. Examples  include systems of entangled atoms and photons~\cite{Haroche_RMP,CohenTBook} and their solid-state analogue~\cite{yamamoto_semiconductor_2000,schoelkopf_wiring_2008,Montambaux_book} or systems of electrons and spins confined in semiconducting nanostructures~\cite{PettaScience05}, to name a few. 

More recently the experimental effort moved toward the realization of hybrid nanosystems obtained interfacing different kinds of quantum information platforms, such as quantum dots, ultracold atoms in cavities, mechanical resonators or superconducting circuits~\cite{MekhovMaschlerRitsch_NatPhys07,MajerNature07,JBlochPRL09,DelbecqetalNatComm12,VerduetalPRL09,ParigiEtAlPRL12,RMP_Esslinger13,MarquardtGirvinArxiv09,
PirkkalainenNature13,KuboetalPRL10,DumeigeEtAlNJP11,Petta_nature12,FreyetalPRB12,SchonPRL12,SchonNJP13,
MiguelSanchezetalNJP13,BergenfeldtArxiv13,ViennotArxiv13,BocquillonEtAlPRL12,DuboisNature13}. 
These novel hybrid architectures are not only of great practical relevance for quantum information but also of fundamental interest as platforms to explore many body physics in novel and unconventional regimes, particularly away from thermal equilibrium. Indeed when small interacting quantum mechanical systems involving inherently different degrees of freedom, such as electrons and photons, are coupled to their fluctuating quantum environments a plethora of interesting many body phenomena are expected to emerge~
\cite{GoldhaberGordon_prl98,Kouwenhoven_science98,Tureci_prl11,Imamoglu_nature11,MebrahtuNature12,JezouinNatComm12,
HofheinzEtAlArxiv11,NazarovChapter,SafiSaleurPRL04,LeHurPRL04,LeHurLiPRB05,LeHurDoucetHofstetterPRL07,
ChungEtAlPRB13,LeClairPhysLett97,LeHur_Kondo_PRB2012,OrthImambekovLeHurPRB13,
GoldsteinetalPRL12,BeraEtAlArxiv13,SbierskiPRL13,Beugnon_Nature06,HenrietEtAl_arxiv14}.
For example, in a recent pioneering experiment~\cite{Delbecq_prl11} the coupling between a quantum dot, a microwave resonator and two biased electronic reservoirs was achieved on chip and signatures of many body correlations observed in the phase of the microwave transmitted photon as a function of the voltage bias across the leads.
Motivated by these rapid developments~\cite{Delbecq_prl11,FreyetalPRB12}, in this paper we discuss the non-equilibrium physics of an hybrid quantum impurity model made by electrons and photons. 

The paper is organized as follows. In section~\ref{sect:model}  we introduce our hybrid quantum impurity model. Section \ref{sect:phase_spect} is devoted to study transmission of photons through the microwave cavity and to elucidate the signature of many body correlations in the phase of the microwave signal. In section~\ref{sect:eff_th} we derive an effective theory for the cavity photon coupled to the density fluctuations of the dot electron, while section~\ref{sect:disc} we present a discussion of the results.

\begin{figure}[t]
\begin{center}
\epsfig{figure=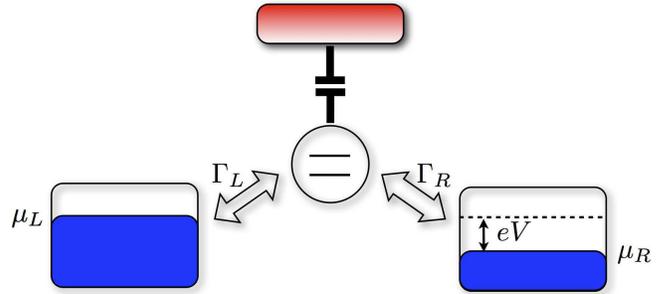,scale=0.29}
\caption{Schematic figure of the hybrid quantum impurity system consisting of a quantum dot hybridized to biased metallic leads and capacitively coupled to an electromagnetic resonator.} 
\label{fig:fig1}
\end{center}
\end{figure}

\section{Hybrid Impurity Model with Electrons and Photons}\label{sect:model}

We model our system as a single level quantum dot tunnel coupled to biased source-drain leads $\alpha=L,R$ and capacitively coupled to a single mode electromagnetic resonator (see Fig.~\ref{fig:fig1}). The Anderson-Holstein type of Hamiltonian for the system reads (setting $\hbar=1$)
\bea\label{eqn:Hsys}
\m{H}_{sys}&=&\sum_{k\sigma\,\alpha}\eps_{k\alpha}c^{\dagger}_{k\alpha\sigma}c_{k\alpha\sigma}+
\sum_{k\sigma\,\alpha}V_{k\alpha}\left(c^{\dagger}_{k\alpha\sigma}d_{\sigma}+h.c.\right)+\nonumber\\&&
+\eps_0\,n+U\,n_{\uparrow}n_{\downarrow}+\lambda\,x\,n+\omega_0\,a^{\dagger}\,a
\eea
where $c_{k\alpha\sigma},d_{\sigma},a$ are annihilation operators of, respectively, leads electrons, correlated dot and photon field, $x=(a+a^{\dagger})/\sqrt{2}$ and $n=\sum_{\sigma}\,d^{\dagger}_{\sigma}\,d_{\sigma}$ while $\eps_{k\alpha}=\eps_k\pm eV/2$. The above Anderson-Holstein model~\cite{HolsteinAnnPhys59,AndersonPRL75} has been studied extensively in the context of non-linear electron transport, with most of the emphasis focused on signatures of many body correlations through electronic properties such as current-voltage characteristics and conductance~\cite{MitraMillis03,CornagliaNessGrempelPRL04,Andrei_PRB12,LaaksoEtAl_NJP14}. Here our scope is to probe the many-body Kondo-type correlations of this system through photon transport.

In order to probe this hybrid system we capacitively couple the resonator to two long transmission lines~\cite{ClerkRMP2010}. Without any loss of generality we consider a symmetric coupling to the two bosonic baths, such that the Hamiltonian of the full system takes the form 
\be\label{eqn:Htot}
\m{H}=\sum_{kl}\,\omega_{k}\,b^{\dagger}_{kl}\,b_{kl}+
\left(a+a^{\dagger}\right)\sum_{kl}\,g_{k}\,\left(b^{\dagger}_{kl}+b_{kl}\right)+\m{H}_{sys}
\ee
where $b_{kl}$ describes photonic modes of the transmission lines ($l=L,R$) with dispersion $\omega_k$ and spectrum $J(\omega)=\pi\sum_k\,g_{k}^2\,\delta(\omega-\omega_k)=2\pi\alpha\omega\,e^{-\omega/\omega_c}$, with $\omega_c$ being an high-energy cut off. In the next section, using exact equations of motion and the input-output relations~\cite{ClerkRMP2010,LeHur_Kondo_PRB2012} we will obtain an expression for the transmission/reflection coefficients which allows to characterize photon transport in the system.

\section{Phase Spectroscopy of a Cavity-Emitted Microwave Photon} 
\label{sect:phase_spect}


\subsection{Input-Output Theory}
Let's start by deriving an input-output relation for this problem, following the treatment of Refs.[\onlinecite{ClerkRMP2010,LeHur_Kondo_PRB2012}]. We consider the full Hamiltonian of the problem $\m{H}$
and we write the equation of motion of the operator $b_{kl}$ which reads
\be
\dot{b}_{kl}=i[H,b_{kl}]=-i\omega_k\,b_{kl}-ig_{k}\,x\,.
\ee
After integration we get for $t>t_0$
\be\label{eqn:bkl_great}
b_{kl}(t)=e^{-i\omega_k(t-t_0)}\,b_{kl}(t_0)-i\,g_{k}\,\int_{t_0}^t\,d\tau\,
e^{-i\omega_k(t-\tau)}\,x(\tau) 
\ee
while if we integrate backward in time from $t_1>t$ we get
\be\label{eqn:bkl_less}
b_{kl}(t)=e^{-i\omega_k(t-t_1)}\,b_{kl}(t_1)-i\,g_{k}\,\int_{t}^{t_1}\,d\tau\,
e^{-i\omega_k(t-\tau)}\,x(\tau)\,.
\ee
We introduce the left input voltage at time $t>t_0$
\be
V_L^{in}(t)
=
\sum_{k}\,g_{k}\,\left(e^{-i\omega_k(t-t_0)}\,b_{kL}(t_0)+hc\right)
\ee 
as well as the output voltage at time $t<t_1$
\be
V_L^{out}(t)
=
\sum_{k}\,g_{k}\,\left(e^{-i\omega_k(t-t_1)}\,b_{kL}(t_1)+hc\right)\,.
\ee 
Using Eq.~(\ref{eqn:bkl_great}-\ref{eqn:bkl_less}) for $b_{kL}(t)$ we can express the average output voltage in terms of the input field. Indeed we have first
\be 
V_L^{out}(t)= V_L^{in}(t)-2\sum_k\,g_{k}^2\,\int_{t_0}^{t_1}\,d\tau\,\sin\left[\omega_k(t-\tau)\right]\,x(\tau)
\ee
then by taking the quantum average from both sides and introducing a susceptibility $\chi_{xx}(t-t')$ defined as
\be
\langle\,x(t)\rangle=\int\,dt'\,\chi_{xx}(t-t')\,\langle\,V^{in}(t')\rangle
\ee
we end up with
\bea
\langle\,V_L^{out}(t)\rangle=\int\,dt''\,r(t-t'')\,\langle\,V_L^{in}(t'')\rangle
\eea
where we have introduced the reflection coefficient
\be
r(t-t')=
\left(\delta(t-t')-i\,\int\,d\tau\,J(t-\tau)
\,\chi^R_{xx}(\tau-t')\right)\,
\ee
with 
$$
J(t)=\sum_k\,g_{k}^2\,\left(e^{-i\omega_k\,t}-e^{i\omega_k\,t}\right)\,.
$$
Using the same approach we can obtain an analog result for the transmission coefficient, $t(\omega)$, defined as
\be\label{eqn:transmission_coeff}
t(\omega)\equiv\frac{\langle V^{out}_R(\omega)\rangle}
{\langle V^{in}_L(\omega)\rangle}=i\,J(\omega)\,\chi^R_{xx}\left(\omega\right)\,.
\ee
%
%

The susceptibility $\chi^R_{xx}$ is nothing but the response of the photon displacement to an input signal, carried by $V^{in}$, and it is given by the Kubo formula
\be
 \chi^R_{xx}(t)=-i\theta(t)\langle\,[x(t),x(0)]\rangle_{H_{sys}}
\ee
where the average is taken over the system hamiltonian with no transmission line but fully including the fermionic leads.
As a result, the expression for the transmission and reflection coefficients have to be interpreted as perturbative in the coupling to the photonic bath ($\alpha\ll 1$). Indeed parameters in figure 2 are chosen such that the scattering matrix is almost unitary, {\it i.e.}, the inelastic contribution to the transmission of light is small compared to the elastic part.

\subsection{Phase Spectroscopy}

Given the previous result for the transmission coefficient, by defining $t(\omega)=\vert\,t\vert\,e^{i\varphi(\omega)}$ we can obtain the phase of the transmitted microwave signal, 
\be\label{eqn:phase}
\mbox{tan}\,\varphi(\omega)\equiv \frac{\mbox{Im}\,t(\omega)}{\mbox{Re}\,t(\omega)}=\frac{\mbox{Re}\,\chi_{xx}^R(\omega)}{\mbox{Im}\,\chi_{xx}^R(\omega)} \,.
\ee

\begin{figure}[t]
\begin{center}
\epsfig{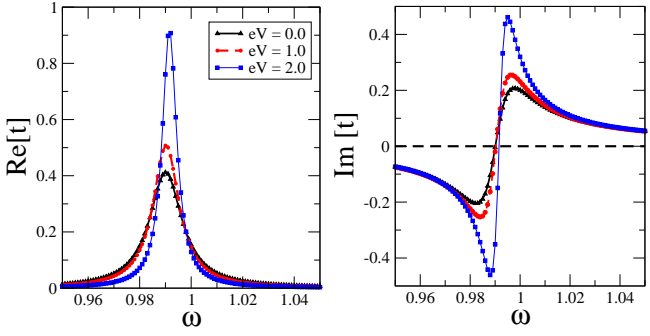}
\caption{Spectroscopy of a transmitted microwave photon for a RLM coupled to a resonator, $U=0$. Real and Imaginary part of the transmission coefficient as a function of the incoming photon frequency for various voltage biases. Units are such that $\Gamma=\omega_0=1$ and $\lambda=0.5\Gamma$. A weak photon dissipation is modeled by a spectral function $J(\omega)=2\pi\alpha\omega\,e^{-\omega/\omega_c}$ with $\alpha=0.001$ and $\omega_c=100\omega_0$.} 
\label{fig:fig2}
\end{center}
\end{figure}
The retarded photon Green's function can be written in Fourier space in terms of the photon self-energy $\Pi^R(\omega)$ as
\be\label{eqn:photon_gf}
\chi^R_{xx}(\omega)=\frac{\omega_0}{\omega^2-\omega_0^2-\omega_0\Pi^R(\omega)}
\ee
where $\Pi^R(\omega)$ includes both the effects of frequency renormalization and the damping due to the fermionic environment.  From this expression we immediately get the frequency dependence of real and imaginary part of the transmission coefficient 
\bea
\mbox{Re}\,t(\omega)=\,\frac{J(\omega)\omega_0^2\,\mbox{Im}\Pi^R(\omega)}
{\left(\omega^2-\omega_0^2-\omega_0\mbox{Re}\Pi^R(\omega)\right)^2+\left(\omega_0\mbox{Im}\Pi^R(\omega)\right)^2}\nonumber\\
\mbox{Im}\,t(\omega)=\frac{J(\omega)\omega_0\left(\omega^2-\omega_0^2-\omega_0\mbox{Re}\Pi^R(\omega)\right)}
{\left(\omega^2-\omega_0^2-\omega_0\mbox{Re}\Pi^R(\omega)\right)^2+\left(\omega_0\mbox{Im}\Pi^R(\omega)\right)^2}\nonumber
\eea
and from them we can extract the phase 
\be
\mbox{tan}\,\varphi(\omega)= \frac{\omega_0\left(\omega^2-\omega_0^2-\omega_0\mbox{Re}\Pi^R(\omega)\right)}
{\,\omega^2_0\,\mbox{Im}\Pi^R(\omega)}\,.
\ee

%

We start evaluating the phase to the lowest order in the electron-photon coupling $\lambda$. To this extent it is convenient to compute directly the photon retarded self-energy $\Pi^R(\omega)$, defined in Eq.~(\ref{eqn:photon_gf}).
Within a Keldysh approach  this turns to be related to the density response function of a purely electronic Anderson Impurity Model (AIM), 
\be 
\Pi^R(t,t')=\Lambda^R(t,t')\equiv \lambda^2\chi_{el}(t-t')
\ee
with $\chi_{el}(t-t')=-i\theta(t-t')\langle[n(t),n(t')]\rangle_{el}$ the electronic charge susceptibility. 
For an Anderson Impurity Model which exhibits a Fermi-Liquid type of ground state this must satisfy the Korringa-Shiba relation~\cite{Filippone_prl11} which implies
\be
\mbox{Im}\chi_{el}(\omega)= \pi\omega\,\left[\left(\mbox{Re}\chi_{el\,\uparrow}(0)\right)^2+
\left(\mbox{Re}\chi_{el\,\downarrow}(0)\right)^2\right]\,.
\ee
However much less is known in general about this quantity at finite frequency in presence of both strong interaction $U$ and finite bias $eV$~(but see references~[\onlinecite{Werner_Keldysh_09,Keldysh_short,DuttAnnals11_1}] for recent developments).

\begin{figure}[t]
\begin{center}
\epsfig{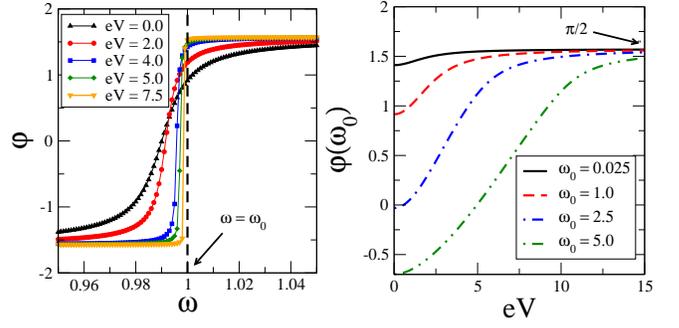}
\caption{Spectroscopy of a transmitted microwave photon for a RLM coupled to a resonator, $U=0$. Phase of the transmission coefficient respectively as a function of the photon frequency (left panel) and on resonance $\omega=\omega_0$ (right panel). In all plots units are such that $\lambda=0.5\Gamma$.} 
\label{fig:fig3}
\end{center}
\end{figure}

\subsection{Resonant Level Model}

To gain some analytical insight we start evaluating the response function for $U=0$. In this non interacting limit, the dot electrons form a resonance of width $\Gamma=\sum_{k\alpha}V_{k\alpha}^2\,\delta\left(\eps_k\right)$ centered at $\eps_0$, their spectral function in the wideband limit becomes a simple lorentzian, and the associated Green's functions read
\be
G^{R/A}_0(\omega)=\frac{1}{\omega-\eps_0\pm i\Gamma}
=\mbox{Re}\,G_0^R(\omega)\mp i\pi\,A_0(\omega)\,
\ee
and
\be
G^<_0(\omega)= \frac{2\pi\,i\,\sum_{\alpha}\,\Gamma_{\alpha}f_{\alpha}(\omega)}{\Gamma}\,A_0(\omega)\,
\ee

Evaluating the density response function reduces then to a convolution of single particle Green's functions
\be
i\Pi^R(\Omega)= \frac{\lambda^2}{2}\,\int\,\frac{d\omega}{2\pi}\,
\left[G^R_0(\omega)G^<_0(\omega-\Omega)+G^A_0(\omega)G^<_0(\omega+\Omega)\right]
\ee
which can be evaluated in closed form. Here we discuss the resulting behavior for the transmission coefficient.
We remind here that we are evaluating the transmission coefficient $t(\omega)$ according to Eq.~(\ref{eqn:transmission_coeff}), which is valid in the linear response regime with respect to the coupling to the photonic bath (external transmission line). 

In figure~\ref{fig:fig2} we plot the behaviour of the real and imaginary part of the transmission coefficient, as well as its phase, as a function of incoming photon frequency $\omega$ and for different bias voltages $eV$. The real part of $t(\omega)$ shows a peak at a renormalized frequency $\omega_{\star}\neq\omega_0$, which shifts and sharpens at larges voltages. Electrons induce a renormalization of photon frequency, which can be seen as a many body light-shift.

The imaginary part of the transmission vanishes at $\omega=\omega_{\star}$ and similarly does the phase $\varphi(\omega_{\star})$, as it follows immediately from Eq.(\ref{eqn:phase}). More interesting and rich is the behavior of the phase at the \emph{bare} resonator frequency, $\varphi(\omega_0)$, which is also what typically measured in experiments~\cite{Delbecq_prl11}.
 As we see from figure~\ref{fig:fig3} at large voltages the phase smoothly approaches $\pi/2$, quite independently from  photon frequency and other system parameters and also from the coupling to the transmission line. This universal behaviour of the phase in the strongly non linear regime can be understood at first by looking at
the structure of photon self-energy $\Pi^R(\omega_0)$.  Indeed one can write the phase of the transmission coefficient on resonance as
\be 
\tan\varphi\left(\omega_0\right)= \frac{\mbox{Re}\,\Pi^R(\omega_0)}
{\mbox{Im}\,\Pi^R(\omega_0)}\,.
\ee

We see how a phase of $\pi/2$ crucially requires the photon damping induced by conduction electrons to vanish sufficiently fast at large bias. Perturbatively we find both the real part and the imaginary part of the self-energy to vanish at voltages larger than a crossover scale $\Lambda(\omega_0)$ as
\be\label{eqn:scalingPi}
\mbox{Re}\,\Pi^{R}(\omega_0)\sim \frac{1}{V^2}\,\qquad\,\mbox{Im}\,\Pi^R(\omega_0)\sim \frac{1}{V^4}
\ee
hence, $\tan\varphi(\omega_0)\sim V^2/\Lambda^2$.  Corrections to the universal large bias limit depend from the photon frequency (bottom right panel), stronger quantum effects (larger $\omega_0/\Gamma$) shift the crossover scale $\Lambda(\omega_0)$ to larger values, suggesting an effective classical picture to explain this effect as we will see later in the paper.

\begin{figure}[t]
\begin{center}
\epsfig{figure=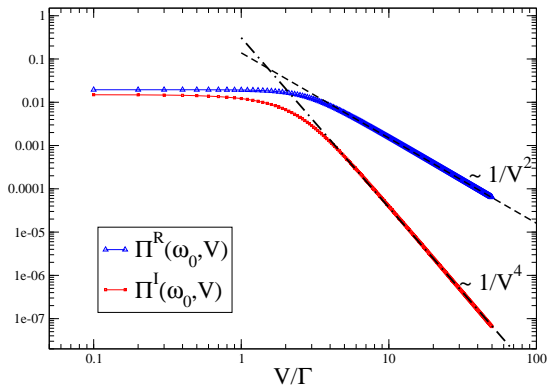,scale=0.3}
\caption{Real and Imaginary part of photon self energy on resonance, $\omega=\omega_0$, as a function of bias voltage. Units are such that $\Gamma=\omega_0=1$ and $\lambda=0.5\Gamma$. At large biases both quantities goes to zero with characteristic power laws.}
\label{fig:fig4}
\end{center}
\end{figure}

\subsection{The Role of Photonic Dissipation}

\begin{figure}[t]
\begin{center}
\epsfig{figure=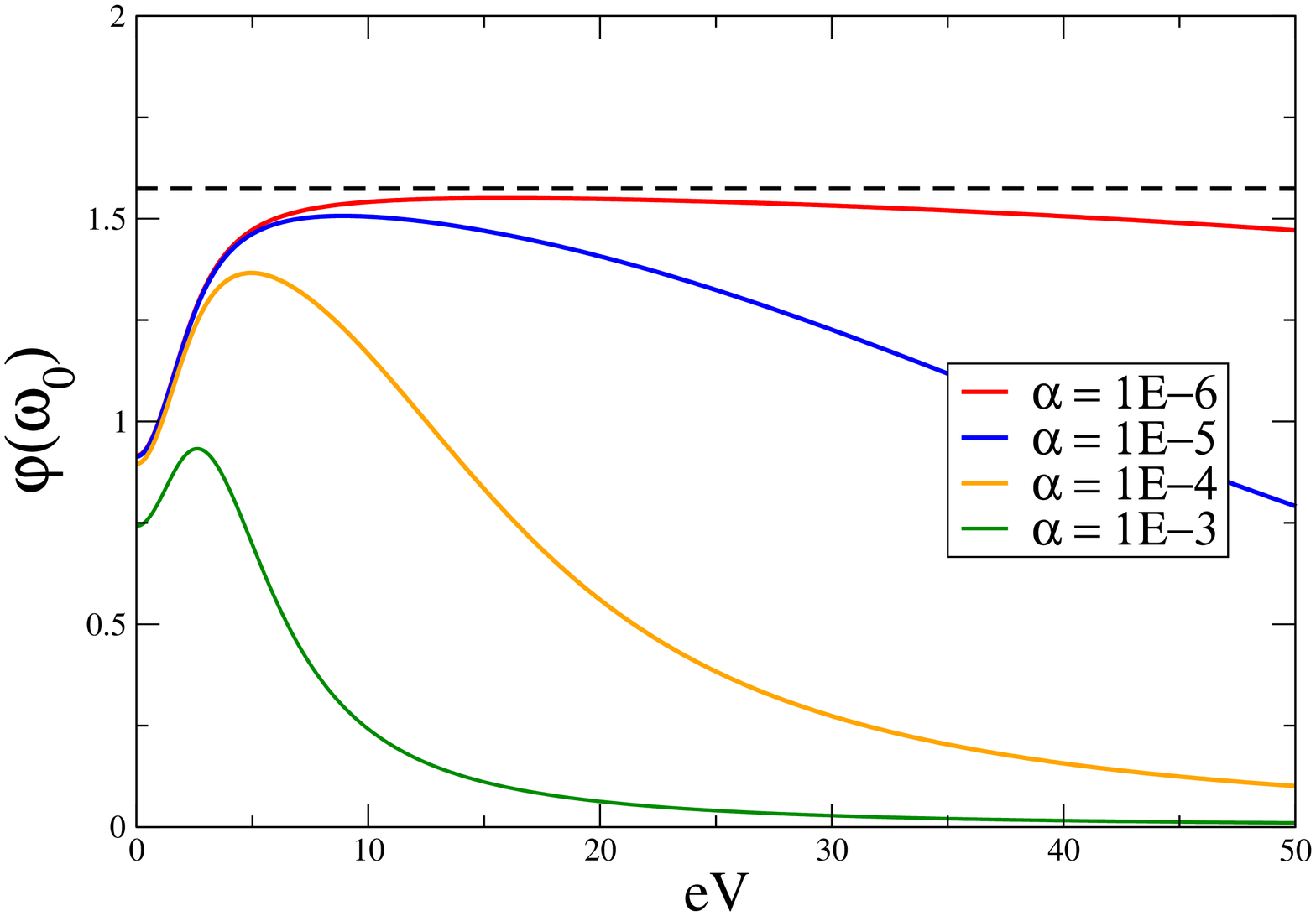,scale=0.3}
\caption{Phase of the transmission coefficient on resonant versus voltage bias in presence of a finite photonic dissipation $\alpha$. Parameters are such that $\Gamma=\omega_0=1$, $\lambda=0.5\Gamma$.}
\label{fig:fig5}
\end{center}
\end{figure}

It is interesting to discuss at this point how the above results change in presence of an extra source of photonic dissipation, such as for example that one coming from a small coupling to the transmission lines. In order to include this effect in our treatment we modify the expression of the (bare) photon Green's function to include some coupling to a dissipative bosonic bath
\be
\chi^{R}_{0xx}(\omega)=\frac{\omega_0}{\omega^2-\omega_0^2+i\omega_0\kappa} 
\ee
where $\kappa=2\pi\alpha\omega_0$ is the effective dissipation induced by a coupling $\alpha$ to the modes $b_{kl}$ in Eq. (\ref{eqn:Htot}). Here we assumed the photonic bath to be Markovian, with a cutoff frequency $\omega_c\gg \omega_0$. In principle, the computation of the photon self-energy to the lowest order is unchanged, since it only involves electronic propagators. As a result we get for the phase of the transmission coefficient the result
\be 
\tan\varphi\left(\omega_0\right)=\frac{\mbox{Re}\Pi^R(\omega_0)}
{\kappa+\mbox{Im}\Pi^R(\omega_0)}\,.
\ee
In figure~\ref{fig:fig5} we plot the phase as a function of the bias voltage for different values of the photonic dissipation. We see that at finite $\kappa\neq0$ the phase crossesover  away from the value $\pi/2$  to $\varphi\rightarrow0$. A simple calculation using the scaling of the photonic self-energy (\ref{eqn:scalingPi}) gives the crossover voltage scale $V_{\star}\sim \kappa^{-1/4}$.

\subsection{Infinite $U$ Anderson Impurity Model}

In this section we briefly consider the $U=\infty$ limit of the Anderson Impurity Model (AIM) and discuss in particular the computation of the density-density response which is relevant to obtain the photon self-energy. In the $U=\infty$ limit we can project out all the states with double occupancy of the fermionic level. This can be done formally by introducing  a representation in terms of an auxiliary fermion and a slave boson,  $d_{\sigma}=f_{\sigma}\,b^{\dagger}$, living in a constrained Hilbert space
\be 
\sum_{\sigma}\,f^{\dagger}_{\sigma}\,f_{\sigma}+b^{\dagger}b=1\,.
\ee
In terms of these new degrees of freedom the AIM Hamiltonian becomes~\cite{Coleman_slaveboson2_PRB87,Read_PRB87}
\be
H_{AIM} = H_0 + \eps_d\,b^{\dagger}b+\sum_{k\alpha\sigma}\left(V_{k\alpha\sigma}\,
c^{\dagger}_{k\alpha\sigma}b^{\dagger}f_{\sigma}+hc\right)
\ee
with $H_0$ the Hamiltonian of a the two free leads.  In order to account for non-equilibrium effects due to the voltage bias we wil use a Keldysh slave boson approach~\cite{ARosch01,RatianiMitraPRB09} and consider the Non-Crossing-Approximation(NCA),
corresponding to the lowest order self-consistent perturbation theory in the hybridization and an exact treatment of the constraint. While in general the NCA equations have to be solved numerically, the limit of large voltage biases has been studied in detail in the context of quantum transport and it is now well understood even at the analytical level~\cite{ARosch01}. Here, due to bias-induced decoherence effects the pseudo fermion acquires a finite damping, $\Gamma_{\star}$, which is given for $\mbox{log}(eV/T_K)\gg1$ by~\cite{ARosch01}
\be 
\Gamma_{\star}\sim \frac{V}{\log^2\left(V/T_K\right)}\left[1+\frac{2}{\log\left(V/T_K\right)}+\ldots\right]
\ee
$T_K$ being the Kondo temperature. This immediately translates into a narrow peak in the pseudo-fermion ($f_{\sigma}$) spectral function, while the slave boson remains incoherent. Since we are interested in computing the electron density response function, which is directly related to the density response of the auxiliary fermion~\cite{Coleman_slaveboson2_PRB87,Read_PRB87}, we can immediately borrow the results obtained previously for the non-interacting case and use them for the $U=\infty$ case as well, with the difference that now the lifetime of the fermionic excitation $\Gamma_{\star}$ is strongly bias dependent. From this we conclude that since for  $\log(eV/T_K)\gg 1$ one has $\Gamma_*\ll eV$, i.e. we are always in the regime where the phase saturates to $\pi/2$.  

\section{Integrating Out the Fermions - Effective Theory for the Photon Field }
\label{sect:eff_th}

To get a further insight, we now derive an effective theory for the photon in the resonator after integrating out all electronic degrees of freedom. We will show that (i) fermionic correlations induce photon non-linearities as well as noise and dissipation and (ii) that the photon long-time dynamics can be described effectively by a Langevin equation for a noisy Duffing oscillator~\cite{DuffingBook}. Finally, we will use this effective theory to interpret the result on the phase spectroscopy obtained in previous section.

In order to proceed  we formulate the problem in terms of Keldysh action~\cite{KamenevBook}.Let's start considering the photon in the resonator that we describe as an harmonic oscillator with coordinate and momentum $x,p$ and with Keldysh action
\be
 \m{S}_{ph}[x(t)] = \int_{\m{K}}\,dt\, \left[\frac{1}{2}\left(\partial_t\,x\right)^2-\frac{\omega_0}{2}x^2\right]
 \ee
where $x(t)$ is a field defined on the two branches Keldysh contour $\m{K}$, or equivalently in the right hand side we have introduced the two components $x_{\pm}$ which are defined on the forward/backward branch, respectively.  It is useful for what will follow to perform a change of variables and introduce the classical and quantum components of the field $x(t)$ defined as
\bea
x_c  = \frac{x_+ + x_-}{2}\\
x_q = \frac{x_+ - x_-}{2}
\eea
in terms of which the action becomes
\be
\m{S}_{ph}[x_c,x_q] = \int\,dt\,2\,x_q\,\left(\ddot{x}_c+\omega_0\,x_c\right)\,.
\ee
A few things are worth to be noticed from this simple result. First, the Keldysh action in this basis vanishes when evaluated at $x_q=0$, i.e.
$$
\m{S}_{ph}[x_c,x_q=0]=0\,.
$$
This is a general result that follow essentially by the unitarity of the evolution and that will hold also for the more complicated problem of electrons coupled to photons we will consider now. Second important observation is that, performing the functional integral over the quantum field we obtain a functional Dirac-delta which constraint the dynamics on the classical trajectory
\be
\ddot{x}_c=-\omega_0\,x_c\,.
\ee
This result is in general true only perturbatively, in other words expanding the Keldysh action to linear order in $x_q$ will gives us the effective quantum dynamics. We will discuss later the meaning of the higher powers of $x_q$.

We now consider the full problem, including the fermionic degrees of freedom, and we couple the photon field to the electronic density so that the full Keldysh action reads
\be 
\m{S}=\m{S}_{ph}[x_c,x_q]+S^{AIM}_{el}[d_{\sigma},d_{\sigma}^{\dagger}]+S_{el-ph}
\ee
where $\m{S}_{ph}$ and $\m{S}^{AIM}_{el}$ are, respectively, the photon and the AIM action while the last term describes electron-photon coupling
\be
 S_{el-ph}=2\lambda\,\int\,dt\,\left[n_q(t)\,x_c(t)+n_c(t)\,x_q(t)\right]
\ee
Here we introduced the so called classical (c) and quantum (q) basis for other operators as well, such the electron density.

We can formally define an effective action for the photon by averaging over electronic degrees of freedom
\be\label{eqn:eff_K_action}
\m{S}_{eff}[x_c(t),x_q(t)]=  \m{S}_{ph}[x_c(t),x_q(t)]+\Upsilon[x_c(t),x_q(t)]
\ee
with 
\be
i\Upsilon =\mbox{log}\langle\,\exp\left(2i\lambda\,\int\,dt\,\left[n_q(t)\,x_c(t)+n_c(t)\,x_q(t)\right]\right)
\rangle_{el}\,.
\ee
This effective theory describes a photon field coupled to the non-equilibrium environment generated by the density fluctuations of dot electrons. While the functional $\Upsilon[x_c(t),x_q(t)]$ cannot be evaluated exactly, due to the form of the electron-photon coupling, the theory admits a simple analytical description in the large bias regime $eV\gg\omega_0$~\cite{HastingsPRB04,NoceraEtAlPRB12}. 

We start noticing that, again, the effective action~(\ref{eqn:eff_K_action}) vanishes when evaluated at $x_q=0$. This is because by construction
\be
e^{i\Upsilon[x_c,x_q=0]}=\langle\,e^{i\int\,dt\,(n_+-n_-)x_c}\rangle_{el}=1 
\ee
the last identity follows from the fact that in absence of any quantum component for the field $x$ we have by construction
\be
\langle\,n_q(t_1)\cdots\,n_q(t_n)\rangle = 0 
\ee
due to causality. Alternatively, this can be seen explicitly from the expression of the effective action. This suggests to expand the effective action in power of the quantum field around $x_q=0$, while keeping for the time being $x_c(t)$ still fluctuating and to stop the theory to the lowest non trivial orders.  By proceeding in this way we obtain
\be
\m{S}_{eff}=\m{S}_{loc}+ \frac{1}{2}\,\int\,dt\int dt'\sum_{\alpha=cl,q}
x_{\alpha}(t)\,\Lambda_{\alpha\beta}(t,t')\,x_{\beta}(t')
\ee
where $\m{S}_{loc}$ is a term which is purely local in time while 
\be
 \Lambda_{\alpha\beta}(t,t') = \langle\,n_{\alpha}(t)n_{\beta}(t')\rangle\,.
\ee
In the next two sections we will discuss separately the physical meaning of both terms. We notice that there are no linear terms nor quadratic terms in the classical field $x_c$ in this expansion, since
\bea
\frac{\delta\m{S}_{eff}}{\delta x_c(t)}\vert_{x_q=0,x_c(t)} = 2\lambda\,\langle\,n_q(t)\rangle_{el} =0\\
\frac{\delta\m{S}_{eff}}{\delta x_c(t)\delta\,x_c(t')}\vert_{x_q=0,x_c(t)} =4i\lambda^2 \langle\,n_q(t)\,n_q(t')\rangle_{el} =0
\eea
again by construction due to the causality of the theory. As we mentioned in the previous paragraph the solution of this non-interacting problem for a generic time dependent field $x_c(t)$ is still a challenging task to be accomplished in fully generality. However we can take advantage of the fact that for very large bias $eV\gg\omega_0$ the dynamics of the electronic degrees of freedom is  much faster than the dynamics of the photon, with typical time scales 
$$
t_{el}\sim 1/eV\ll t_{ph}\sim\,1/\omega_0\,.
$$
As a result we can to leading order assume an adiabatic approximation and assume that the electronic degrees of freedom just see a static classical field $x_c$. In other words we can evaluate the coefficient of the effective action by considering a AIM which is in its instantaneous ground state corresponding to the configuration of the classical field $x_c\equiv x_c(t)$.

\subsection{Electron Induced Photon Non-Linearities}

The first term is local in time and reads
\be 
\m{S}_{loc}=\int\,dt\,2\,x_q\,\left(\ddot{x}_c+\omega_0^2\,x_c+F(x_c)\right)
\ee
with $F(x_c)=\lambda\,\langle n_c\rangle_{el}$. We notice that, in absence of other terms in $\m{S}_{eff}$,
the functional integral over $x_q$ would result in the photon equation of motion $
 \ddot{x}_c+\omega_0^2\,x_c+F(x_c)=0$
from which we can interpret $V_{eff}(x)=\frac{\omega_0^2\,x^2}{2}+\int_0^x\,dx'\,F(x') $
as an effective potential for the photon, that renormalizes the bare harmonic one. Expanding it around $x=0$ we get
the general form
\be\label{eqn:Veff}
V_{eff}(x)= \frac{\lambda\,x}{2}+\frac{\omega_*^2\,x^2}{2}+\eta\,x^3+g\,x^4
\ee
where $\omega_*,\eta,g$ represent respectively the renormalized frequency, the anisotropy term and the effective
anharmonicity.  We can obtain analytical expressions for these coefficients in terms of the parameters of the electronic problem, in particular the external gate and bias voltages.
Their expressions can be obtained by differentiating $V_{eff}(x)$ in Eq.~(\ref{eqn:Veff}) and read for $U=0$, respectively
\be\label{eqn:freq}
\omega_*^2=\omega_0^2-\frac{\lambda^2}{\pi}\,\sum_{\alpha}\,\frac{\Gamma_{\alpha}}{\left(\eps_0-\mu_{\alpha}\right)^2+\Gamma^2} 
\ee
and
\be\label{eqn:anisotropy}
\eta = \frac{2\lambda^3}{\pi}\,\sum_{\alpha}\,\frac{\Gamma_{\alpha}\left(\eps_0-\mu_{\alpha}\right)}{\left(\eps_0-\mu_{\alpha}\right)^2+\Gamma^2} 
\ee
and finally the anharmonicity
\be\label{eqn:anharm}
g = \frac{2\lambda^4\,\Gamma_{\alpha}}{\pi}\,\sum_{\alpha}\,
\frac{\Gamma^2-\left(\eps_0-\mu_{\alpha}\right)^2}{\left[\Gamma^2+\left(\eps_0-\mu_{\alpha}\right)^2
\right]^3}\,.
\ee

The anisotropy term $\eta$ is particularly interesting to discuss. As we see from Eq.~(\ref{eqn:anisotropy}) this term at zero gate voltage $\eps_0=0$ reduces to 
\be
\eta(\eps_0=0) = -\frac{2\lambda^3}{\pi}\,\sum_{\alpha}\,\frac{\Gamma_{\alpha}\,\mu_{\alpha}}{V^2/4+\Gamma^2} 
\ee
namely it vanishes for symmetric bias voltages, $\mu_L+\mu_R=0$. This is indeed not surprising since the anisotropy term  breaks the inversion symmetry $x\rightarrow\,-x$, which is a direct consequence of particle hole symmetry of the Holstein model that holds at $\eps_0=0$ and $\mu_L=-\mu_R$. The results discussed here concern the non-interacting $U=0$, but can be extended to the interacting $U=\infty$ case in the large bias regime, as discussed in previous section upon substituting $\Gamma\rightarrow\Gamma_{\star}$ in Eqns.~(\ref{eqn:freq},\ref{eqn:anisotropy},\ref{eqn:anharm})

\begin{figure}[t]
\begin{center}
\epsfig{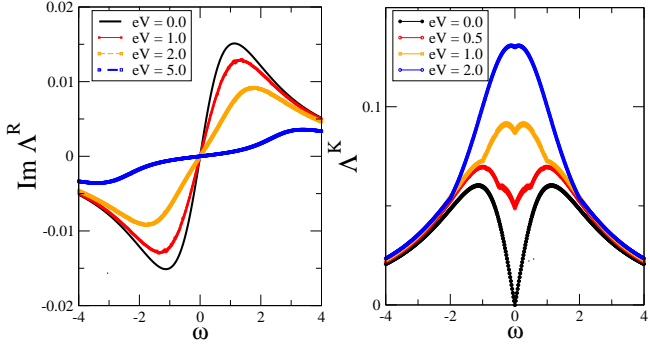}
\caption{Effective theory for the photon. Retarded $\mbox{Im}\Lambda^R(\omega)$ and Keldysh $\Lambda^K(\omega)$ component of the fermionic-induced environment (here for $U=0$) giving rise to damping and dissipation for the photon.  In all plots (except the right bottom) units are such that $\Gamma=\omega_0=1$, $\lambda=0.5\Gamma$.} 
\label{fig:fig6}
\end{center}
\end{figure}
\subsection{Electron Induced Photon Damping and Dissipation}

Let's now consider the effects of the other terms in the effective action which are non local in time. We can distinguish two contributions,  $\m{S}_{frict}$ and $\m{S}_{noise}$, which include retardation effects and read, respectively
\bea
\m{S}_{frict}=
\int\,dt\,dt' x_{c}(t)\,\mbox{Im}\,\Lambda_{x_c}^R(t-t')\,x_{q}(t')\\
\m{S}_{noise}=
\int\,dt\,dt' x_{q}(t)\,\Lambda_{x_c}^K(t-t')\,x_{q}(t')
\eea
with $\Lambda^{R,K}_{x_c}(t-t')=2i\lambda^2\langle [n(t)\,,n(t')]_{\pm}\rangle_{el}$ being respectively the retarded (R) and Keldysh (K) component of the density-density response function of the AIM. We now focus on the low frequency behaviour of the kernels $\Lambda^R_{x_c}(\omega),\Lambda^K_{x_c}(\omega)$. 
 
We start from the retarded component, encoding the spectrum of density fluctuations. At low frequency, $\omega\rightarrow0$,  its imaginary part vanishes linearly, see figure~\ref{fig:fig6}
\be\label{eqn:gamma}
\mbox{Im}\,\Lambda^R(\omega)=-i\gamma(x_c)\,\omega
\ee
where the low frequency slope can be computed analytically for the RLM case, $U=0$ and reads
\be\label{eqn:gamma_rlm}
\gamma(x_c)=\frac{2\,\lambda^2\,\Gamma}{\pi^2}\,\sum_{\alpha}\,
\frac{\Gamma_{\alpha}}{\left[\left(\mu_{\alpha}-\eps_0-\lambda\,x_c\right)^2+\Gamma^2\right]^2}\,.
\ee
More generally, this result can be also understood by noticing that, upon integrating out the driven cavity, the problem becomes reminiscent of an AC driven quantum dot coupled to fermionic leads~\cite{ButtikerPhysLettA93,GabelliScience06,FeveScience07,GabelliRepProgPhys2012} which has been recently shown to possess an effectively ohmic dynamical charge response~\cite{CMora_natphys,NiggLopezButtiker_PRL96,ButtikerNiggEPJB09,EtzioniHorovitzLeDoussalPRL11,HamamotoEtAlPRB10,Filippone_prl11,FilipponePRB13,
DuttPRB13}. A finite bias voltage does not change qualitatively this linear behavior and only renormalize slightly the slope. 

As opposite, the Keldysh component is strongly affected by a finite bias voltage which turns the characteristic linear in frequency equilibrium behaviour, imposed by fluctuation-dissipation theorem at $T=0$, into a finite value at zero frequency (see Fig.~\ref{fig:fig6}). We can obtain a full analytical understanding of this in the case of a biased RLM.  Using the expression for the lesser and greater RLM Green's function we can write the Keldysh component  as
\bea\label{eqn:lambda_keld}
\Lambda^K(\omega;x_c)&=&i\,\frac{8\pi\lambda^2}{\Gamma^2}\,\sum_{\alpha\beta}\,
\Gamma_{\alpha}\Gamma_{\beta}\,\int\,d\Omega\,A_0(\Omega)\times\\
&\times &
\left[A_0(\Omega+\omega)\,f_{\alpha}(\Omega)f_{\beta}(-\Omega-\omega)+ 
\left(\omega\leftrightarrow-\omega\right)
\right]\nonumber\,.
\eea
The low frequency behaviour of this kernel controls the nature of the noise. It is easy to see already from this expression that for $\Omega\rightarrow0$ the behaviour of $\Lambda^K$ depends strongly from the bias voltage. While at zero bias the kernel vanishes as a power law, see fig~\ref{fig:fig2}, a finite voltage acts as an infrared cutoff and results into a finite value at zero frequency, $\Lambda^K(\omega\rightarrow 0;x_c)\equiv iD(x_c)$, with
\be
iD(x_c)=i\,\frac{8\pi\lambda^2}{\Gamma^2}\,\Gamma_L\,\Gamma_R
\int_{-eV/2}^{eV/2}\,d\omega\,A_0(\omega)\,A_0(\omega)
\ee
which plays the role of an effective diffusion coefficient $D$ for the dynamics of the photon.  A simple calculation gives 
\bea \label{eqn:diffusion}
D
=\frac{4\lambda^2\,\Gamma_L\Gamma_R}{\pi\Gamma^3}\sum_{a=\pm}\,
\arctan\left(\frac{eV/2-a\eps_0-a\lambda\,x_c}{\Gamma}\right)+\nonumber\\
+\frac{4\lambda^2\,\Gamma_L\Gamma_R}{\pi\Gamma^3}\sum_{a=\pm}
\frac{\Gamma\left(eV/2-a\eps_0-a\lambda\,x_c\right)}{\left(eV/2-a\eps_0-a\lambda\,x_c\right)^2+\Gamma^2}
\nonumber\,.
\eea
Quite interestingly  we see that the diffusion coefficient vanishes linearly with the bias voltage as $eV\rightarrow0$
\be
D(x_c;V)\sim \left(\frac{8\lambda^2\Gamma_L\Gamma_R}{\pi\Gamma^2\left(\left(\eps_0+\lambda\,x_c\right)^2+\Gamma^2\right)}\right)\,eV
\ee
which suggests in some sense that the bias voltage is playing the role of an effective temperature, $T_{eff}\sim eV$. For larger values of the applied bias the diffusion coefficient saturates to a constant value, independent on the oscillator position $x_c$, which reads
\be
D(x_c)\sim  \frac{4\lambda^2\Gamma_L\Gamma_R}{\Gamma^3}\,.
\ee
Again as in the previous section, one can extend the above discussion to the interacting $U=\infty$ case in the large bias regime upon substituting $\Gamma\rightarrow\Gamma_{\star}$ in the above results.

Finally, if we translate the above results back in the time domain we can write the effective action as
\be 
\m{S}_{eff}=\m{S}_{loc}+\int\,dt\gamma(x_c)\,x_q\,\dot{x}_c+\frac{i}{2}\,\int\,dt\,D(x_c)x^2_q(t)\,.
\ee
This immediately shows that the retared component of the electron density fluctuation gives rise to a friction term in the photon effective action with damping coefficient $\gamma(x_c)$, while the Keldysh term plays the role of a noise source with effective diffusion coefficient $D(x_c)$. To make explicit the connection with a classical stochastic Langevin dynamics we introduce the noise field $\xi(t)$ through an Hubbard-Stratonovich decoupling~\cite{KamenevBook}
and then perform the integral over the quantum component which gives the classical dynamics
\be
\ddot{x}_c=-\omega_0\,x_c-F(x_c)-\gamma(x_c)\dot{x}_c+\xi(t) 
\ee
 with $\langle \xi(t)\,\xi(t')\rangle=D(x_c)\delta(t-t')$ a white multiplicative noise.  If now one couples this system to a coherent mode of the photonic transmission line playing the role of an explicit driving term, see Eq.(\ref{eqn:Htot}),  and with the force field $F(x)$ deriving from the effective anharmonic potential in Eq.~(\ref{eqn:Veff}),  the above dynamics describes a Duffing oscillator~\cite{DuffingBook}  driven by noise~\cite{MallickEJPB04}, which has recently attracted interest in the context of circuit QED~\cite{LarsonArxiv13,PeanoPRA12} and nanomechanical systems~\cite{DykmanPRB12}.

\subsection{Effective Temperature, Photon Occupation and Dissipated Electronic Power}

We conclude this analysis by discussing in this section the nature of the environment generated by fermionic density fluctuations.
In order to do so we start recalling that for a generic bosonic Green's function $g(\Omega)$ in thermal equilibrium the following fluctuation dissipation relation (FDR) holds between the Keldysh and the retarded/advanced components 
\be
g^K(\Omega)=\coth{\frac{\beta\Omega}{2}}\,\left(g^R(\Omega)-g^A(\Omega)\right) 
\ee
where $\beta=1/T$ is the temperature of the system. In a generic out of equilibrium condition this relation does not hold, however it is still interesting to define an effective distribution function
\be\label{eqn:F_function}
f(\Omega)=\frac{g^K(\Omega)}{g^R(\Omega)-g^A(\Omega)} 
\ee
and to discuss its frequency dependence. In particular, a singular behaviour at low frequency $\Omega\rightarrow0$
\be\label{eqn:Teff}
F(\Omega\rightarrow0) = 2T_{eff}/\Omega 
\ee 
it is often used to define an \emph{effective temperature} for the out of equilibrium system, a concept that has emerged in many different contexts both classical and quantum~\cite{MitraMillisPRB05,MitraEtAlPRL06,FoiniCugliandoloGambassi_PRB11,DallaTorreetalPRB2012,RibeiroSiKirchner_EPL13,SchiroMitraArxiv13}. In the case of interest here the photon resonator is coupled to the effective bath induced by the electronic density fluctuations, encoded in the functions $\Lambda^K(\omega),\Lambda^R(\omega)$. We define therefore a photon effective distribution function as
\be 
F(\omega)=\Lambda^K(\omega)/\mbox{Im}\,\Lambda^R(\omega)\,.
\ee
\begin{figure}[t]
\begin{center}
\epsfig{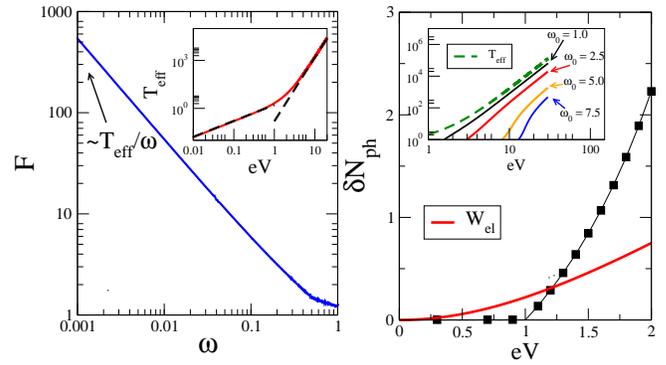}
\caption{Effective theory for the photon. Low frequency behavior of the distribution function and voltage induced effective temperature (left), giving rise to an effective thermal occupancy for the photon at large bias (right, inset), compared to the electronic dissipated power $W_{el}=I_{el}V$. In all plots (except the right bottom) units are such that $\Gamma=\omega_0=1$, $\lambda=0.5\Gamma$} 
\label{fig:fig7}
\end{center}
\end{figure}
We plot in figure~\ref{fig:fig7} the behaviour of this distribution function.
It is easy to check analytically that, at zero bias $eV=0$, the bosonic self-energy satisfies the FDR, i.e.
\be
\Lambda^K(\Omega)= \coth{\frac{\beta\Omega}{2}}\,\left(\Lambda^R(\Omega)-\Lambda^A(\Omega)\right)\,\qquad\,\qquad\,eV=0
\ee
However, at finite bias a characteristic bosonic $1/\omega$ singularity emerges at low frequency, that allows to define a bias-dependent effective temperature $T_{eff}$.  In particular, using the definition~(\ref{eqn:Teff}) and the results for the diffusion coefficient and the damping
\bea
\Lambda^K(\Omega\rightarrow0)=iD(V)\\
\mbox{Im}\Lambda^R(\Omega\rightarrow0)=i\gamma(V)\Omega
\eea
we obtain an effective equilibrium result for the effective temperature
\be
T_{eff}(V)=\frac{D(V)}{4\gamma(V)}
\ee
that we plot in the inset of figure~\ref{fig:fig7} as a function of the bias voltage. We see the small and large bias behaviours (compared to the electronic lifetime $\Gamma$)  are characterized by two different power laws, $T_{eff}\sim V$ at small bias when $T_{eff}$ is almost set by the noise $D(V)\sim V$ while $T_{eff}\sim V^4$ at large voltage when the noise as we have seen saturates while the dissipation decays fast $\gamma(V)\sim 1/V^4$.

Quite interestingly, not only the bath distribution function but also the photon occupation $\langle N_{ph}\rangle$ is effectively thermal at large voltages (see right panel). In order to clarify this point we now compute the photon
occupation as a function of the bias voltage. Let's start considering the photon occupation, $N_{ph}=\langle\,a^{\dagger}\,a\rangle$. This can be obtained from the Keldysh component of the photon Green's function as
\be
 1+2\,N_{ph}=\langle\,x^2\rangle=\frac{i}{2\pi}\,\int\,d\omega\,\chi^K(\omega)\,.
\ee
In absence of electron-photon interaction, $\lambda=0$, we have
\be
\chi_0^K(\omega)=-\pi\,i\,\left( \delta(\omega-\omega_0)-\delta(\omega+\omega_0) \right) \coth\frac{\omega}{2T}
\ee
from which we immediately get the thermal occupation $N_{ph}=\left(\exp\left(\omega_0/T\right)-1\right)^{-1}$.

In presence of a finite interaction $\lambda\neq0$ the expression for the Keldysh component generally reads
\be\label{eqn:keldysh}
\chi^K(\Omega)=\chi^R(\Omega)\left[
\left(\chi_0^R\right)^{-1}\,\chi_0^K(\Omega)\,\left(\chi_0^A\right)^{-1}+\Pi^K(\Omega)
\right]\,\chi^A(\Omega)
\ee
where the retarded/advanced component reads
\be
\chi^{R,A}(\Omega)=\frac{1}{\left(\chi_0^{R,A}\right)^{-1}-\Pi^{R,A}(\Omega)}
\ee
while $\Pi^{R,A,K}$ are respectively the retarded/advanced/keldysh component of the photon self-energy, which we computed to lowest order in perturbation theory in $\lambda$ as already discussed in previous sections.

It is now easy to see that for finite $\lambda$ the first term in Eq.~(\ref{eqn:keldysh}) gives a vanishing contribution to the photon occupation. Therefore, for the sake of computing the photon occupation we may write
\be
 1+2\,N_{ph}=\frac{i}{2\pi}\,\int\,d\omega\,\chi^R(\omega)\,\Pi^K(\omega)\,\chi^A(\omega)\,.
\ee
For weak coupling the retarded photon Green's function is significantly different from zero only close to $\omega=\pm\omega_0$ (neglecting small frequency renormalization effects), while the Keldysh component of the self-energy varies smoothly around $\omega=\pm\omega_0$.  This suggests to write the retarded and advanced photon Green's function as
\be
\chi^R(\omega)= \frac{\omega_0}{\left(\omega^2-\omega_0^2\right)+i\omega_0\,\Pi^R_{im}(\omega_0)}
\ee
and to evaluate the above integral to get
\be 
  1+2\,N_{ph}\sim \frac{\Pi^K(\omega_0)}{\Pi^R_{im}(\omega_0)}\,.
\ee
In figure~\ref{fig:fig7} we plot the behaviour of the the photon number as a function of the voltage bias respectively for weak and large biases and for different photon frequencies $\omega_0$. We see that the small bias regime features a clear threshold behavior, in accordance with previous results~\cite{MitraMillis03} namely the photon occupation sets in only as $V\sim \omega_0$. This has to be understood in terms of energetic balance to excite a photon in the resonator by means of conduction electrons with energy $eV$ above the 
Fermi sea. As opposite, the large bias regime shows a steep increase of the photon number, with a power-law behaviour that can be understood in term of an effective temperature description. Indeed as we have seen $\Pi^K(\omega_0)$ can be identified with the noise (or diffusion coefficient) $D(V)$ while $\Pi^R_{im}(\omega_0)\sim\gamma(V)\,\omega_0$ so that we have
\be
N_{ph}\sim \frac{D(V)}{\gamma(V)\omega_0}\sim \frac{T_{eff}}{\omega_0}\sim V^4 
\ee
as we discussed in the previous section. Finally, we would like to clarify the relation between the photon occupation as a function of the bias voltage and the dissipated electronic power. In order to do this we consider the weak coupling case and proceed by perturbative evaluation of the photon and electron Green's functions.
It is tempting to compare the photon dissipated power $W_{ph}=\omega_0\,N_{ph}$ with the electronic power due to the current flow, that generally reads $W_{el}=IV$. However this identification reveals a rather different behaviour both at small and at large voltage biases. Indeed for vanishing bias $eV\rightarrow0$ we expect
\be
W_{el}\sim G_0\,V^2 \,\qquad\,V\ll\Gamma
\ee
 while for large biases, when the current saturates we would expect
 \be
W_{el}\sim I_{*}\,V \,\qquad\,V\gg\Gamma \,.
 \ee
  These different power laws, that should survive also in presence of electron-photon coupling are difficult to reconcile with the result for the photon occupation.  We finally remark that  at higher frequencies departures from this effective equilibrium appear and the resulting semiclassical stochastic dynamics in this non-equilibrium environment~\cite{BraddeBiroliArxiv12} deserves further investigations.

\section{Discussion}\label{sect:disc}

We can finally interpret the phase spectroscopy in light of the effective photon model valid at large bias. In this regime the non-linearity is only subleading  and the main effects due to fermions are a many-body light frequency shift and a damping,
so that one can write the photon propagator as
\be
\chi^R_{xx}(\omega) = \frac{1}{\omega^2-\omega_*^2-i\omega_0\gamma} 
\ee
resembling the spin susceptibility of an Ohmic spin-boson model for $\omega\sim\omega_\star$~\cite{LeHur_Kondo_PRB2012}. This renormarlized frequency $\omega_{\star}$ can be identified with a many-body Lamb shift. The crucial observation is now that, quite differently from conventional regimes, a large bias pushes the system into the regime $\vert\omega_0-\omega_*\vert\gg\gamma$, from which a phase of $\pi/2$ immediately follows on resonance. 

We stress that this result is sensitive to external sources of decoherence for the photonic system, such as for example those provided by the auxiliary transmission lines. Indeed the presence of a finite dissipation $\kappa$ would contribute to the photon damping with a finite value even in the large bias regime and then diminish the phase of the transmitted photon. As a result one should expect the phase to reach a maximum close to $\pi/2$ and then to cross over to zero for sufficiently large voltage biases, $V\gg V_{\star}(\kappa)$. Yet, for small values of the photon dissipation we expect a sizeable window of voltages where the phase stays close to $\pi/2$, as confirmed by a direct evaluation of the transmission coefficient in presence of dissipation.

\section{Conclusions} In this paper we discussed the physics of an hybrid quantum impurity model with interacting electrons and photons. We have shown that the phase of the transmitted photon at large voltage biases approaches the value of $\pi/2$ and we interpreted this result in light of an effective photon theory obtained integrating out the electronic degrees of freedom. It would be interesting to explore the connection between this result and the electronic (Friedel) phase shift at low bias, related to the Kondo effect. Finally we have discussed the robustness of the above result with respect to photon dissipation and shown that this hybrid system can be used as a simulator of the stochastic Duffing oscillator in relation with chaos.

\section{Acknowledgment} It is a pleasure to thank M. B\" uttiker, A. Houck, T. Kontos, A. Browaeys, L. Henriet, C. Mora, J. Petta, P. Dutt, M. Filippone, L. Messio, O. Parcollet, Z. Ristivojevic, P. Simon, J. Gabelli, J. Esteve,  and H. Tureci  for interesting discussions and comments related to this work. This work has been funded by the LABEX PALM Paris-Saclay, grant Quantum Dyna and by DOE  through DE-FG02-08ER46541 (KLH).


%
%
%


\end{document}